\documentclass[12pt]{article}
\usepackage{authblk}
\usepackage{hyperref}
\usepackage[top=25truemm,bottom=28truemm,left=24truemm,right=24truemm]{geometry}
\usepackage{setspace}
\normalsize
\usepackage{amsmath,amssymb,mathrsfs,amsbsy,latexsym,amsfonts,amsthm}
\usepackage{fancyhdr}
\pdfoutput=1 % For pdfLaTeX
\usepackage{graphicx}
\usepackage{color}
\usepackage{comment}
\usepackage{here}
\usepackage{braket}
\usepackage[utf8]{inputenc}
\usepackage{mathtools}

\usepackage{bbm}

\newcommand{\cH}{\mathcal{H}}

\DeclareMathOperator{\tr}{tr}

\usepackage{epsf}
\newcommand{\cO}{\mathcal{O}}
\newcommand{\cA}{\mathcal{A}}

\numberwithin{equation}{section}
\title{\bf %Remarks on 
Bulk Reconstruction and Gauge Invariance
}

\author[1]{
Sotaro~Sugishita\thanks{\tt sotaro(at)gauge.scphys.kyoto-u.ac.jp}
}
\author[2]{
	Seiji~Terashima\thanks{\tt terasima(at)yukawa.kyoto-u.ac.jp}
}
\vspace{5mm}
\affil[1]{\it\normalsize Department of Physics, Kyoto University, Kyoto 606-8502, Japan}
\affil[2]{\it\normalsize 
Center for Gravitational Physics and Quantum Information,  
\mbox{Yukawa Institute for Theoretical Physics, Kyoto University, Kyoto 606-8502, Japan}  }
\date{}
\begin{document}

\maketitle
\thispagestyle{fancy}
\renewcommand{\headrulewidth}{0pt}

%%%%%%%%%%%%%%%%%%%%%%%%%%%%%%%%%%%%%%%%%%%%%%%%%%%%%%%%%%%%%%
%%%%%%%%%%%%%%%%%%%%%%%%%%%%%%%%%%%%%%%%%%%%%%%%%%%%%%%%%%%%%%
\begin{abstract}

In this paper, we discuss the concept of bulk reconstruction, which involves mapping bulk operators into CFT operators to understand the emergence of spacetime and gravity. We argue that the $N=\infty$ approximation fails to capture crucial aspects of gravity, as it does not respect gauge invariance and lacks direct connections between energy and boundary metrics. Key concepts such as entanglement wedge reconstruction and holographic error correction codes, which are based on the  $N=\infty$ theory, may be incorrect or require significant revision when finite $N$ effects are considered. We present explicit examples demonstrating discrepancies in bulk reconstructions and suggest that a gauge-invariant approach is necessary for an accurate understanding.

\end{abstract}

\newpage
\thispagestyle{empty}
\setcounter{tocdepth}{2}

\setlength{\abovedisplayskip}{12pt}
\setlength{\belowdisplayskip}{12pt}

\tableofcontents
\newpage
%%%%%%%%%%%%%%%%%%%%%%%%%%%%%%%%%%%%%%%%%%%%%%%%%%%%%%%%%%%%%%
%%%%%%%%%%%%%%%%%%%%%%%%%%%%%%%%%%%%%%%%%%%%%%%%%%%%%%%%%%%%%% 

\section{Introduction and summary}
%%%%%%%%%%%%%%%%%%%%%%%%%%%%%%%%%%%%%%%%%%%%%%%%%%%%%%%%%%%%%%
%%%%%%%%%%%%%%%%%%%%%%%%%%%%%%%%%%%%%%%%%%%%%%%%%%%%%%%%%%%%%%

The AdS/CFT correspondence \cite{Maldacena:1997re} can be regarded as a definition of quantum gravity.
To achieve this, we need to consider a finite $N$ (holographic) CFT because the $N=\infty$ limit of CFT is not a well-defined conformal theory.
Then we will study a large $N$ (asymptotic) expansion of physical quantities to interpret the CFT as a bulk gravitational theory.
For such an interpretation, 
it is important to understand how to represent bulk operators as CFT operators \cite{Banks:1998dd, Balasubramanian:1998sn, Bena:1999jv, Hamilton:2006az, Kabat:2012hp, Kabat:2013wga, Kabat:2015swa, Kabat:2011rz, Terashima:2021klf}.
This procedure is called the bulk reconstruction.
In other words, it provides an answer to the question of how spacetime and gravity emerge.

The holographic CFT$_d$ for
$N=\infty$ is expected to be
the $d$-dimensional generalized free field (GFF) \cite{Duetsch:2002hc},
which is just the $d+1$ dimensional 
free theory where the radial direction is regarded as an internal space of the Kaluza-Klein theory.
This $d+1$ dimensional free theory
corresponds to the free bulk theory.
Here, we emphasize that this free gravity theory does not impose the ``Gauss law'' constraints of the gauge invariance (diffeomorphism invariance) of the gravity theory and the energy is not related to the metric of the asymptotic boundary \cite{Donnelly:2015hta, Donnelly:2016rvo}.
Thus, the $N=\infty$ approximation misses the most important aspect of gravity and is crucially different from the finite $N$ theory.

On the other hand, some important notions in the bulk reconstruction 
are based on $N=\infty$ theory, such as the entanglement wedge reconstruction, subregion duality, and the holographic error correction code.
Furthermore, these concepts are expected to be valid even when we include $1/N$ corrections. At least, it is generally expected that there are no significant differences between the $N=\infty$ theory and the finite $N$ theory.
These notions are based on the importance of considering subregions of spacetime; for example, the horizon and black hole are related to the concept of subregions.
However, given the crucial differences between the $N=\infty$ theory and the finite $N$ theory discussed above, 
a serious reconsideration of these expectations is required.
Furthermore, it has been claimed that these notions in bulk reconstruction are shown to be invalid or significantly modified, based on explicit computations in the AdS/CFT correspondence, as demonstrated in \cite{Terashima:2020uqu, Terashima:2021klf, Sugishita:2022ldv, Terashima:2023mcr, Sugishita:2023wjm}. 
The differences between the $N=\infty$ theory and the finite $N$ theory are crucial for these results.\footnote{
Our basic standpoint is that we want to understand finite \( N \) CFTs. (When \( N = \infty \), CFTs are not well-defined.)  
When \( N \) is large, physical quantities can sometimes be approximated well by \( 1/N \) expansion.
Even in cases where this approximation is valid, it may hold only perturbatively, or it may extend to capture nonperturbative effects as well.
Although sometimes nonperturbative effects are called finite \( N \) effects, in this paper, by finite \( N \) we mean working with finite-\( N \) CFTs rather than the \( N = \infty \) limit.  
The reason we emphasize finite \( N \) is that there is a significant difference in subregion bulk reconstruction between finite-\( N \) CFTs and free bulk theories at \( N = \infty \) (generalized free theories). 
In this paper, what we are actually discussing is that problems already arise at the perturbative level in the 
\( 1/N \) expansion.
}

In this paper, 
we argue that 
the reason why such widely accepted properties of bulk reconstruction should be either incorrect or significantly modified is, indeed, that they do not respect the gauge invariance in the interacting gravitational theory corresponding to the finite $N$ theory.
We will consider the simplest case: the (global) vacuum state, which corresponds to the pure global AdS space, and take a ball-shaped subregion whose entanglement wedge corresponds to the AdS-Rindler wedge. Even for this simplest case, the discussion is non-trivial, and we have the differences between the $N=\infty$ theory and the finite $N$ theory.
Indeed, we will provide explicit examples demonstrating that global and AdS-Rindler bulk reconstructions yield different operators, and that the original entanglement wedge reconstruction is not valid.
These also demonstrate that 
the subregion complementarity \cite{Sugishita:2023wjm},
rather than the holographic error correction code, is realized to solve the radial locality ``paradox'' in \cite{Almheiri:2014lwa}.
It should be emphasized here that all such discussions, especially the argument that the properties of the holographic error correction code do not exist in the holographic CFT, are presented entirely in the language of CFT.

Then, we will explain that this property is naturally understood within a gauge-invariant bulk description.
It is important to note that there are no gauge invariant local operators in the gravitational theory,
and the gravitational dressing is needed to make naive local operators gauge invariant \cite{Donnelly:2015hta, Donnelly:2016rvo}.
There are discussions on the ambiguities associated with gravitational dressing, for example, in \cite{Chen:2019hdv, Akers:2019wxj, Bao:2019hwq, Bahiru:2022oas}.

%XXXX
%Let us consider the subregion $A$ which is the right half of the space on the $t=0$ slice and 
%the Rindler patch $D(A)$.
%The  Minkowski space is divided into the right Rindler wedge $D(A)$, left Rindler wedge $D(\bar{A})$, the future wedge, and the past wedge.

\subsection{Summary of our claims}

Below, we summarize our claims on the several ``established'' properties of AdS/CFT. 
The explanation of the details of our claims 
will be described in later sections.

\paragraph{Quantum error correction codes in AdS/CFT}
The bulk reconstruction claims that bulk operators (which are low-energy operators) can be reconstructed by CFT operators in multiple ways. 
These CFT operators are distinct in the entire CFT Hilbert space, including high-energy states. 
However, 
the holographic quantum error correction (QEC) proposal claims that these different operators are the same in the low-energy subspace (code subspace) as stated in \cite{Pastawski:2015qua}.
If we take $N=\infty$ where the CFT is the generalized free field (GFF) theory, which is equivalent to the free bulk theory, 
it can be regarded as QEC as proposed in \cite{Almheiri:2014lwa}. 
However, the code subspace is equivalent to the entire Hilbert space of GFF because the GFF only contains the low-energy modes,
i.e. it is trivial as a quantum error correction code.\footnote{
One might think that for $N=\infty$ the CFT Hilbert space will be decomposed into GFFs around
the semi-classical backgrounds, like the vacuum or black holes, and
the GFF for the vacuum is the code subspace.
However, for $N=\infty$ the GFF is completely decoupled from 
other sectors.
}
Thus, for the holographic QEC proposal, it is crucial to see whether this structure still holds including $1/N$ corrections.
Indeed, if we include the leading order correction in the $1/N$ expansion to the GFF,
which is the non-vanishing leading order of the three-point functions in CFT, 
bulk fields constructed from CFT operators on different subregions
becomes {\it different in code subspace} in general.
Therefore, the quantum error correction code proposal works only
for $N=\infty$ rather trivially.
For finite $N$, the relevant question for the holographic QEC proposal is whether 
the two bulk operators, which are equivalent for $N=\infty$, with $1/N$ corrections can be equivalent in the code (low energy) subspace
with the interactions in the bulk theory.
The answer is no.
(If some properties of the free bulk theory are completely different from those of the interacting bulk theory, the free theory is not useful for them.)

\paragraph{Entanglement wedge reconstruction}

Let us consider a bulk local operator supported on the intersection of the entanglement wedges of two different CFT subregions $A$ and $B$ for $N=\infty$. 
We can reconstruct the bulk operator as the CFT operators $\cO_A$ and $\cO_B$ (which are supported on $A$ and $B$ respectively) for $N=\infty$. 
We now consider $1/N$ corrections to them.
Then we can show that
these two CFT operators $\cO_A$ and $\cO_B$ cannot be the same at the leading order correction in $1/N$ expansion in general,
even around the vacuum which corresponds to the pure AdS spacetime. 
Thus, the entanglement wedge reconstruction \cite{Dong:2016eik} which claims $\cO_A = \cO_B$ does not work including $1/N$ corrections.
On the other hand, a weaker version of the entanglement wedge reconstruction, stated in \cite{Terashima:2020uqu, Sugishita:2022ldv, Sugishita:2023wjm}, may be valid.
This claims that parts of (smeared) bulk local operators with $1/N$ corrections in the entanglement wedge of a CFT subregion can be reconstructed from the CFT operators supported on the subregion, while the ones outside the entanglement wedge cannot be reconstructed. The set of reconstructable bulk operators is smaller than that of naive local operators in the usual entanglement wedge reconstruction.

\paragraph{Subregion duality from relative entropy }
The relative entropies in the bulk and the CFT for the subregion may be the same up to ${\cal O}(1/N)$ \cite{Jafferis:2015del} and this seems to lead to the subregion duality and the entanglement wedge reconstruction as in \cite{Dong:2016eik}. 
However, this should not be valid at ${\cal O}(1/N)$\footnote{
This is the first non-trivial 
leading order of the three-point function 
which is needed for the results of  \cite{Jafferis:2015del}.
}
as shown in \cite{Sugishita:2022ldv}.
This may be because in \cite{Dong:2016eik} it was assumed that the Hilbert space for the bulk gravitational theory is factorized for the subregions. The assumption of the factorization is (approximately) valid only if we fix the gauge.
The correct understanding of the bulk relative entropy in \cite{Jafferis:2015del} will be one using the algebraic entanglement entropy
where the bulk subalgebra is generated by the gauge invariant operators supported on the entanglement wedge,
and it leads to a subregion duality and an entanglement wedge reconstruction in the sense discussed in \cite{Harlow:2016vwg} \cite{Cotler:2017erl}. 
We claim that subregion duality and the entanglement wedge reconstruction based on the algebraic entanglement entropy are completely different from the usual ones
although it appears to be assumed that there is not much significant difference between them in \cite{Harlow:2016vwg} \cite{Cotler:2017erl}.
In particular, the weaker version of the entanglement wedge reconstruction \cite{Terashima:2020uqu, Sugishita:2022ldv, Sugishita:2023wjm} is related to the gauge invariant operators on the bulk region as we argue.\footnote{
Including the $1/N$ corrections, the bulk local operators need to attach the gravitational dressing \cite{Donnelly:2015hta, Donnelly:2016rvo}, which is the critical reason for the non-factorization of the Hilbert space.
This gravitational dressing is related to the weaker version of the entanglement wedge reconstruction. 
}

The main points that have been stated above are as follows:
There are gaps between the GFF and finite $N$ case or $1/N$ corrected one concerning the consideration of subregions.
Note that the GFF and the free bulk theory are equivalent, 
and thus this may be regarded as the discrepancy between the bulk effective theory and quantum gravity (=CFT).

\section{Bulk reconstruction with gauge fixing}
In this section, we review the global and AdS-Rindler (HKLL) bulk reconstruction \cite{Hamilton:2006az} by taking care of $N=\infty$ or finite $N$. The distinction between $N=\infty$ and finite $N$ is essential to see the difference between the global and AdS-Rindler reconstruction. 
The discussion will be done in a gauge-fixed way as in the standard argument of the bulk reconstruction \cite{Hamilton:2006az}. 
The difference between the two reconstructions will be obvious in a gauge-invariant argument as we will see in section~\ref{sec:gauge-inv}.

\subsection{Global AdS}
We consider the $d$-dimensional holographic CFT on the cylinder with coordinates $(\tau, \Omega)$. The corresponding bulk is the $(d+1)$-dimensional global AdS and we take the bulk metric as 
\begin{align}
    ds^2=\frac{1}{\cos^2\! \rho}\left(-d\tau^2+d \rho^2+ \sin^2 \!\rho\, d\Omega_{}^2\right).
\end{align}
First, we consider $N=\infty$ case,
i.e. bulk free theory in global AdS or the GFF theory on the boundary.
We can solve the EOM of the bulk free theory by the mode expansion
for the global AdS coordinates and
obtain the creation operators $a^\dagger_{n l m}$.
Then, the bulk local field $\phi$ can be 
expanded by these modes.
In particular, the boundary value of $\phi$, which is given by 
\begin{align}
 O(\tau, \Omega)=\lim_{z \to 0} z^{-\Delta} \phi(\tau, z, \Omega) 
\end{align}
where $z=\cos (\rho)$ and $\Delta$ is the conformal dimension of the CFT operator corresponding to $\phi$.
Here, \(\phi\) is expressed as a linear combination of $a_{n l m}$ and $a^\dagger_{n l m}$ as in the usual field theory on Minkowski spacetime. 
Conversely, we can express $a^\dagger_{n l m}$ by the boundary values $O(\tau, \Omega)$.
Then, inserting this into the mode expansion of $\phi$, we have 
the bulk reconstruction formula from the boundary values:
\begin{align}
    \phi (\tau,\rho,\Omega) = \int d \tau' d\Omega' K(\tau,\rho, \Omega;\tau',\Omega')  {\cal O}(\tau',\Omega'),
\end{align}
where $K(\tau,\rho, \Omega;\tau',\Omega')$ is a specific function called the smearing function 
whose explicit form was given in \cite{Hamilton:2006az} and called the HKLL bulk reconstruction formula.
It is important to note that the  two-point function 
$
\langle 0 | O(\tau, \Omega)  O(\tau', \Omega') | 0 \rangle$
computed in the free bulk theory coincides with
the two-point function of the primary operator with the conformal dimension $\Delta$ of CFT, which is universal, up to a numerical normalization factor.
If we regard this $O(\tau, \Omega) $ as a $d$-dimensional ``CFT'' operator,  the ``CFT'' is a generalized free field theory \cite{Duetsch:2002hc}.

If we define 
\begin{align}
    \phi^G (\tau,\rho,\Omega) \equiv \int d \tau' d\Omega' K(\tau,\rho, \Omega;\tau',\Omega')  {\cal O}^{CFT} (\tau',\Omega'),
    \label{gHKLL}
\end{align}
for a primary operator ${\cal O}^{CFT} (\tau,\Omega)$ of any finite $N$ CFT,
the bulk two-point function 
is reproduced by the CFT operator $\phi^G $:
\begin{align}
\langle 0 | \phi^G (\tau,\rho,\Omega) \phi^G (\tau',\rho',\Omega') | 0 \rangle    
=\langle 0 | \phi (\tau,\rho,\Omega) \phi (\tau',\rho',\Omega') | 0 \rangle, 
\label{e1}
\end{align}
bacause ${\cal O}$ and ${\cal O}^{CFT}$ have the same two-point functions.\footnote{
There were some discussions on AdS$_2$ case \cite{Dey:2021vke}.
}
In particular, an $N=\infty$ limit of the holographic CFT around the vacuum is expected to be this GFF because of the large $N$ factorization.
Then, \eqref{gHKLL} in this limit reproduces the bulk $n$-point function and gives the HKLL bulk reconstruction formula.

Note that the smearing function  
$K(\tau,\rho, \Omega;\tau',\Omega') $ has ambiguities although there is no ambiguity for expressing $\cO$ by the creation operators.
Indeed, for $N=\infty$, 
if we replace it as
$  K(\tau,\rho, \Omega;\tau',\Omega') \rightarrow  K(\tau,\rho, \Omega;\tau',\Omega') +\delta K(\tau,\rho, \Omega;\tau',\Omega')  $,
the reconstructed bulk operator  
$\phi^G (\tau,\rho,\Omega)$ is invariant if 
\begin{align}
\int d \Omega' \int_{-\infty}^{\infty} d\tau' \, e^{i \omega \tau} Y_{lm}(\Omega) \, \delta K(\tau,\rho, \Omega;\tau',\Omega') =0,
\nonumber
\end{align}
is satisfied for
$
\omega= 2n+l+\Delta 
$
where $n$ is a non-negative integer although $\omega$ can take any real number as a Fourier transformation of $\tau$.

\paragraph{$1/N$ corrections}
The HKLL bulk reconstruction can be extended 
to include $1/N$ corrections, which correspond to the interaction in the bulk 
by requiring non-linear EOM or a kind of micro causality \cite{Kabat:2012hp, Kabat:2013wga, Kabat:2015swa},\footnote{
The requirements were described in the holographic gauge. However, the HKLL bulk reconstruction may not give a local operator in the holographic gauge as discussed in \cite{Donnelly:2015hta}.
} 
which is given as
\begin{align}
\phi^{G(1)} (\tau,\rho,\Omega) & \equiv  \int d \tau' d\Omega' K(\tau,\rho, \Omega;\tau',\Omega')  {\cal O}^{CFT} (\tau',\Omega') \nonumber \\
& + \frac{1}{N}\int  d \tau' d\Omega' d \tau'' d\Omega'' 
 K^{(1)ab}(\tau,\rho, \Omega;\tau',\Omega', \tau'',\Omega'')  {\cal O}^{CFT}_a (\tau',\Omega') {\cal O}^{CFT}_b (\tau'',\Omega''),
 \label{cor1}
\end{align}
where ${\cal O}^{CFT}_a$ is a low-energy primary operator
and $K^{(1)ab}$ is $N$-independent.
It is noted that if we take $\phi$ a spherical symmetric operator, i.e. it is $\Omega$-independent, 
each term of $1/N$ expansion is a spherical symmetric operator.

\subsection{AdS-Rindler bulk reconstruction}

Next, we review the AdS-Rindler bulk reconstruction \cite{Hamilton:2006az}.
Let us consider the AdS- Rindler patch of $AdS_{d+1}$ with the metric 
\begin{align}
\label{AdSRind-metric}
    ds^2=-\xi^2 dt_R^2+\frac{d\xi^2}{1+\xi^2}+(1+\xi^2) d \chi^2,
\end{align}
where $\chi$ stands for the coordinates of the $d-1$ dimensional hyperbolic space.
Its asymptotic boundary on the $t_R=0$ slice, which will be denoted by $A$, is a ball-shaped subregion in the sphere, whose entanglement wedge $M_A$ is the AdS-Rindler patch itself.
For this, we can repeat the bulk reconstruction for the global AdS: we solve the EOM to obtain the mode expansion and
rewrite the bulk local operator $\phi$ by the modes.
Then, the bulk local operator can be expressed by the boundary values of the bulk local operator using a smearing function $K^R$\footnote{
Precisely speaking, 
the smearing function $K^R$ does not exist for the bulk local operator. 
We always need to consider the smeared bulk operator (distribution) \cite{Morrison:2014jha}.
It is also noted that $K^R$ is unique.}
Using the smearing function, we define  
\begin{align}
 \phi^R (t_R,\xi,\chi) \equiv \int dt'_R d\chi' K^R(t_R,\xi,\chi; \chi',t'_R)  {\cal O}^{CFT}(\chi',t'_R).
 \label{phiR}
\end{align}
This CFT operator $\phi^R$ reproduces the bulk two-point function correctly even for finite $N$ as
\begin{align}
    \langle 0 | \phi (X) \phi (X') | 0 \rangle
=
\langle 0 | \phi^R (X) \phi^R (X') | 0 \rangle,
\label{e2}
\end{align}
for arbitrary two points in $M_A$,  
where $\ket{0}$ is the global vacuum (not the Rindler vacuum).
Furthermore, they satisfy 
\begin{align}
     \langle 0 | \phi (X') \phi (X) | 0 \rangle
=\langle 0 | \phi^G (X') \phi^G (X) | 0 \rangle=
\langle 0 | \phi^R (X') \phi^G (X) | 0 \rangle,
\label{e3}
\end{align}
for arbitrary $X$ in the entire bulk and $X'$ in the entanglement wedge $M_A$.
For $N=\infty$, we can also reproduce any higher-point bulk correlation function in the entanglement wedge $M_A$ as the global reconstruction does. 
Thus, the bulk local operators in the entanglement wedge $M_A$ 
are reconstructed from the CFT operators supported on the subregion $A$ for $N=\infty$.

Note that for $N=\infty$,
the CFT is regarded as the GFF, which is just the free bulk theory on AdS.
Thus, $\phi^G$ and $\phi^R$ are the same operators.\footnote{
This is possible by the ambiguities of $K$, i.e. we may have $K^R=K+\delta K$.
}
In particular, the creation and annihilation operators are related by a unitary transformation (i.e., the Bogoliubov transformation with operators on $M_{\bar{A}}$).
This means that the holographic error correction code in the $N=\infty$ theory,
the map (for example, the Petz map) is trivial.
This is because the code subspace (low energy subspace) and the physical space are the same for $N=\infty$.

Furthermore, if we include even the first non-trivial corrections in the $1/N$ expansion, the structure of the holographic error correction code is lost \cite{Sugishita:2023wjm}
as we will see later.
Here, we will give some remarks on it.
Let us consider the difference between the two operators $\phi^G$ in \eqref{gHKLL} and $\phi^R$ in \eqref{phiR}:
\begin{align}
    \phi^\delta \equiv \phi^R-\phi^G,
\end{align}
in the finite $N$ CFT, i.e., we use the HKLL formulas for $\phi^G$ in \eqref{gHKLL} and $\phi^R$ in \eqref{phiR} with ${\cal O}^{CFT}$ for finite $N$ CFT although the smearing functions $K$ and $K^R$ are those for $N=\infty$ theories without $1/N$ corrections.
This $\phi^\delta$ satisfies, for arbitrary point $X$ in the entire bulk, 
\begin{align}
   \bra{0} \phi^\delta \, {\cal O}_a(X) \ket{0}=0,
   \label{2p}
\end{align}
where ${\cal O}_a(X)$ is any primary operator of the CFT,
because this is a two-point function between the primary operators corresponding to $\phi$ and ${\cal O}_a(X)$ which vanishes by \eqref{e3}.
This exactly implies 
\begin{align}
     \phi^\delta \ket{0}=0.
\end{align}
However, it does not imply $\phi^\delta =0$ and we may have $\phi^\delta \ket{\psi} \neq 0$ for some $\ket{\psi}$.
In particular, some three-point functions  can be nonzero:
\begin{align}
   \bra{0} {\cal O}_a (X) \phi^\delta \, {\cal O}_b(X) \ket{0} \neq 0.
\end{align}

We indeed have such an example for CFT on $d=2$ Minkowski spacetime $(t,x)$. 
Using the lightcone coordinates $u=t-x$, $v=t+x$,
let us consider 
\begin{align}
    \tilde{\phi}^\delta:= \int du dv \, e^{iu p_u +i v p_v} {\cal O}_\Delta (u,v),
\end{align}
for a non-chiral scalar primary operator ${\cal O}_\Delta$ with $p_u p_v <0$ which implies the energy $p^t$ is lower than the absolute value of the momentum $p^x$.
Then, we find
\begin{align}
   \bra{0} \tilde{\phi}^\delta \, {\cal O}_\Delta (u',v') \ket{0}= \int du dv \, e^{iu p_u +i v p_v} (u-u'+i \epsilon)^{-\Delta}  (v-v'+i \epsilon)^{-\Delta}=0,
\end{align}
where $\epsilon>0$ is due to the ordering of the operators \cite{Terashima:2023mcr}
because if $p_u <0$ the integration path $u \in \mathbf{R}$ will be deformed to $u \in \mathbf{R}-i \infty$ 
without crossing a singular point.
On the other hand, for the three-point function with the energy-momentum tensor, we have
\begin{align}
   & \bra{0} T(u'') \tilde{\phi}^\delta \, {\cal O}_\Delta (u',v') \ket{0} \nonumber \\
   = & \int du dv \, e^{iu p_u +i v p_v}
    \frac{\Delta}{2} 
      \frac{(u-u'+i \epsilon)^2}{(u''-u+i \epsilon)^2(u''-u'+2 i \epsilon)^2}
    \frac{1}{(u-u'+i \epsilon)^{ \Delta} (v-v'+i \epsilon)^{ \Delta} }
   \neq 0,
\end{align}
because there are singular points for $u$ in the integrand both in 
the upper and lower half-plane. 
Thus, we have $\tilde{\phi}^\delta \ket{0}=0$ and $\tilde{\phi}^\delta \neq 0$.

Like this example, $\phi^G$ and $\phi^R=\phi^G+\phi^\delta$ can be distinguished for finite $N$ CFT by looking at three-point functions.
The difference $\phi^\delta$ comes from the $1/N$ correction, and then we can ask
whether we can take $\phi^{G(1)}=\phi^{R(1)}$
including the $1/N$ correction, like \eqref{cor1}, at this order.
In other words, the question that we should ask is whether $\phi^\delta$ can be canceled by the CFT operator supported only on $A$ by order by order in $1/N$ expansions, and the answer is no as shown in \cite{Sugishita:2022ldv} by giving an explicit example. 
We will show it again in the next section with a refined example.

\section{Subregion complementarity and bulk reconstruction}

In this section, we will show first that 
the global and AdS-Rindler bulk reconstructions should give different operators if we include $1/N$ corrections.
On the other hand, both of the bulk theories on the global and the AdS-Rindler patches may be consistent.
Indeed, these two theories have 
the same operators on 
an overlapped region (which is the AdS-Rindler patch) of these two patches
for the free bulk limit (i.e. $N=\infty$),
while the operators will be different for finite $N$.
Here, it should be emphasized that all results in this section are shown entirely in the language of CFT.
We called the existence of the two different descriptions depending on the observers, the subregion complementarity \cite{Sugishita:2023wjm}. 
We will give an example that illustrates what the subregion complementarity is.
In the next section, we will explain that
the subregion complementarity results from the intrinsic non-locality of the operators in the gravitational theory due to the gauge invariance.

\subsection{Global and AdS-Rindler bulk reconstructions give different operators}
\label{differentop}

Let us consider the CFT on $S^{d-1}$ and take a ball-shaped subregion $A$ which is slightly larger than half of the $S^{d-1}$. 
We then consider a spherically symmetric smeared bulk local operator $\tilde{\phi}$ in the global AdS such that $\tilde{\phi}$ is supported only on the entanglement wedge $M_A$ associated with the CFT subregion $A$.
We also require $\tilde{\phi}^\dagger=\tilde{\phi}$.
By the global HKLL bulk reconstruction (or another bulk reconstruction that maintains the symmetry), we have 
$\tilde{\phi}_G \equiv \int_{S^{d-1}} K(X) {\cal O}^{CFT}(X)$ with $\tilde{\phi}_G^\dagger =\tilde{\phi}_G$ where $K(X)$ is a real and spherically symmetric function.
Below, we will use the CFT energy-momentum tensor $T_{00}(\Omega)$ on the $\tau=0$ slice and the Hamiltonian $H=\int_{S^{d-1}} d \Omega \, T_{00}(\Omega)$. 
Here, $T_{00}(\Omega)$ is shifted by a constant from the usual definition in CFT so that we have $\bra{0}H \ket{0}=\bra{0}T_{00}(\Omega) \ket{0}=0$, which implies $H \ket{0}=0$.
Then, by the rotational symmetry, we have
\begin{align}
      \bra{0}[\tilde{\phi}_G , [T_{00}(\Omega) ,\tilde{\phi}_G ]] \ket{0} 
     =\frac{1}{V(S^{d-1})} \bra{0}[\tilde{\phi}_G , [H ,\tilde{\phi}_G ]] \ket{0} 
    =  \frac{2}{V(S^{d-1})} \bra{0} \tilde{\phi}_G H \tilde{\phi}_G  \ket{0} >0,
\end{align}
for any $\Omega$.
This implies that $\bra{0}[\tilde{\phi} , [T_{00}(\Omega) ,\tilde{\phi} ]] \ket{0} ={\cal O}(N^0)$.\footnote{
Note that $T_{\mu \nu}  \sim N h_{\mu \nu}$ where the ``graviton'' operator $h_{\mu \nu}$ is normalized 
such that it has the conventional normalization of the CFT two-point function, which is independent of $N$. In the large $N$ expansion, we use $h_{\mu \nu}$, then the three-point function is ${\cal O}(1/N)$. 
}

On the other hand, 
since $\tilde{\phi}$ is localized on $M_A$, we can reconstruct it by the AdS-Rindler HKLL reconstruction as $\tilde{\phi}^R$.
Then, for $\Omega \in \bar{A}$, we will have
\begin{align}
     \bra{0}[\tilde{\phi}^R , [T_{00}(\Omega) ,\tilde{\phi}^R ]] \ket{0}=0,
\end{align}
by the micro causality (i.e., any operators on $A$ commutes with $T_{00}(\Omega)$ for $\Omega \in \bar{A}$). Note that this discussion uses the low-energy states only.\footnote{
This implies that $(\tilde{\phi}^R)^2 \ket{0} \neq (\tilde{\phi})^2 \ket{0}$ although  $\tilde{\phi}^R \ket{0} = \tilde{\phi} \ket{0}$ as discussed in the previous subsection.
On the other hand, using the Reeh–Schlieder theorem, we have an operator $\varphi^R$ supported on $A$ such that $\varphi^R \ket{0} =(\tilde{\phi})^2 \ket{0}$, but it is impossible that
$\varphi^R = (\tilde{\phi}^R)^2 $ even for the low-energy states.}
Any $1/N$ corrections to $\tilde{\phi}$ cannot contribute to $\cO(N^0)$ terms in $\bra{0}[\tilde{\phi} , [T_{00}(\Omega) ,\tilde{\phi} ]] \ket{0}$. 
Therefore,
the AdS-Rindler bulk reconstruction of a bulk operator in the entanglement wedge and the global bulk reconstruction of the same bulk operator cannot be the same CFT operators, even if we include $1/N$ corrections appropriately.\footnote{
Here, we define $\tilde{\phi}^R$ as an operator on the entire $S^{d-1}$, since this is the form adopted in the discussions of~\cite{Dong:2016eik, Almheiri:2014lwa, Pastawski:2015qua}. Furthermore, we demonstrate in Appendix \ref{reducedtoA} that even when restricted to a subregion $A$, the two operators do not agree.
}

%\footnote{
%More precisely, these two bulk operators, which are supposedly the same in the bulk theory, are different because of the gauge fixings. 
%}

We can also see the difference by considering the energy density of the following state: 
\begin{align}
\label{psiG-e}
    \ket{\psi}= (1+i \epsilon \tilde{\phi}_G+ \frac{1}{2} (i \epsilon \tilde{\phi}_G )^2) \ket{0},
\end{align}
where $\epsilon$ is a small real parameter.
By the rotational symmetry, we have
\begin{align}
    \bra{\psi}  T_{00}(\Omega) \ket{\psi} = \frac{1}{V(S^{d-1})} \bra{\psi}  H \ket{\psi}
    =  \epsilon^2 \frac{1}{V(S^{d-1})} \bra{0} \tilde{\phi}_G H \tilde{\phi}_G  \ket{0}+ {\cal O}(\epsilon^3),
\end{align}
where $V(S^{d-1})$ is the volume of the time slice $S^{d-1}$.
Thus, $\bra{\psi}  T_{00}(\Omega) \ket{\psi}$ is ${\cal O}(\epsilon^2)$ for any point $\Omega$.
On the other hand, using $\tilde{\phi}_R$ instead of $\tilde{\phi}_G$, we consider 
\begin{align}
    \ket{\psi^R}= (1+i \epsilon \tilde{\phi}^R+ \frac{1}{2} (i \epsilon \tilde{\phi}^R )^2) \ket{0}.
\end{align}
If we take $\Omega \in \bar{A}$, we have
\begin{align}
    \bra{\psi^R}  T_{00}(\Omega) \ket{\psi^R} & =    
    \frac{1}{2} \epsilon^2 \bra{0}[\tilde{\psi} , [T_{00}(\Omega) ,\tilde{\phi}^R ]] \ket{0} \nonumber \\
    &   \;\;\; -\frac{1}{2} i \epsilon^3 \bra{0} ( (\tilde{\phi}^R)^2 T_{00}(\Omega) \tilde{\phi}^R - \tilde{\phi}^R T_{00}(\Omega) (\tilde{\phi}^R)^2 )\ket{0}
    +{\cal O}(\epsilon^4)\nonumber \\
    &= -\frac{1}{2} i \epsilon^3 \bra{0} ((\tilde{\phi}^R)^2 T_{00}(\Omega) \tilde{\phi}^R - \tilde{\phi}^R T_{00}(\Omega) (\tilde{\phi}^R)^2 )\ket{0}
    +{\cal O}(\epsilon^4),
\end{align}
where we have used the micro causality $[T_{00}(\Omega) ,\tilde{\phi}^R ]=0$ for $\Omega \in \bar{A}$.
Thus, $\bra{\psi^R}  T_{00}(\Omega) \ket{\psi^R}$ is ${\cal O}(\epsilon^3)$.
Therefore, we have $\bra{\psi}  T_{00}(\Omega) \ket{\psi} \neq \bra{\psi^R}  T_{00}(\Omega) \ket{\psi^R}$, and $\ket{\psi^R}$ are different from $\ket{\psi}$.

\subsection{Some illustrative examples}
%%%%%%%%
%%%%%%%%
\begin{figure}[htbp]
%\vspace{-0.01\columnwidth}
\centering
\includegraphics[width=15em]{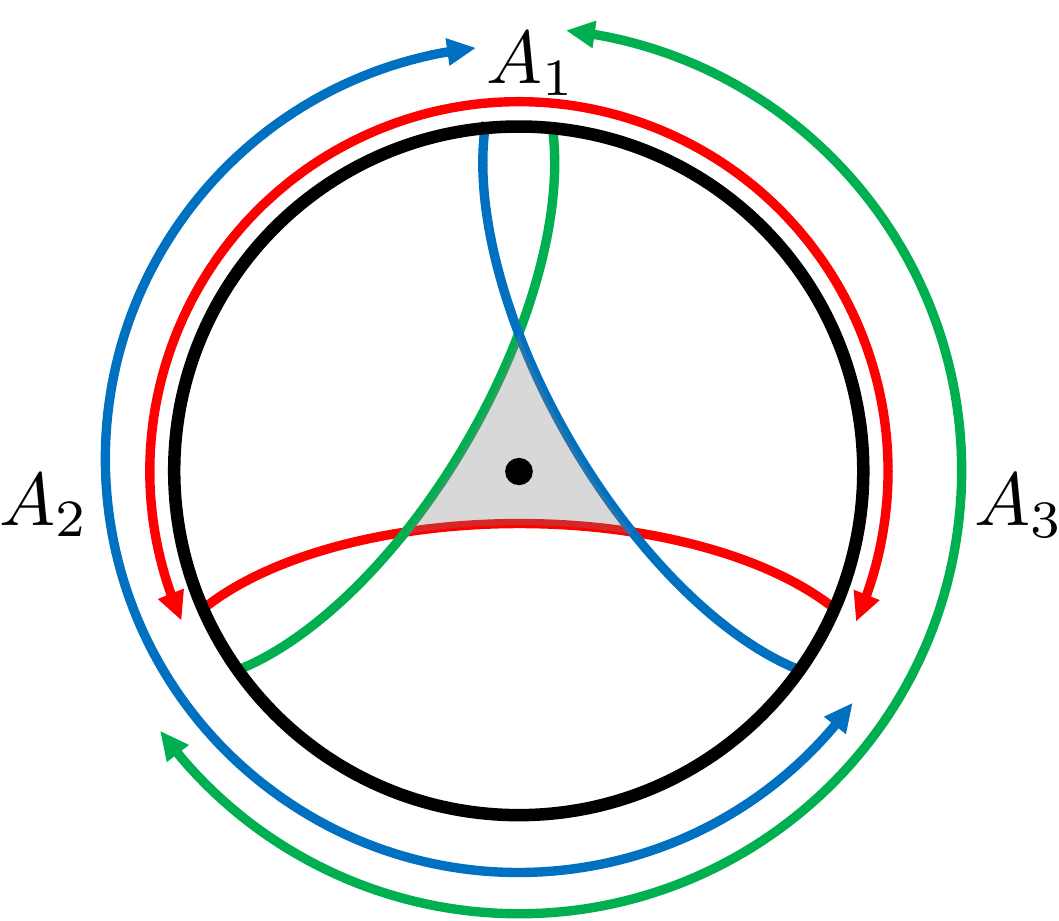}
\vspace{-0.01\columnwidth}
\caption{We take three CFT subregions $A_1, A_2, A_3$ on a time slice. The bulk subregion (with gray color) around the center is the intersection of the time-slices of the entanglement wedges associated with $A_i$. The bulk subregion does not touch the boundary, and we call it an island-like region.}
\label{fig:island}
\vspace{-0.01\columnwidth}
\end{figure}
%%%%%%%%
%%%%%%%%

Let us consider an example that exhibits the relation between the subregion complementarity and gauge invariant operators.
The setup is similar to the discussion of the bulk locality in \cite{Almheiri:2014lwa}, although the conclusions are completely different.
We consider a time slice of global $AdS_3$ spacetime. 
The corresponding 
CFT is on $S^1$ parameterized by $ 0 \leq \theta < 2\pi$.  
We take a subregion $A_1$ of the CFT by $0 \leq \theta < \theta_1$ where $\theta_1$ satisfies $\pi < \theta_1 < \frac{4 \pi}{3}$.
We also take the subregions $A_2$ and $A_3$ 
by the $\frac{2 \pi}{3}$ and $\frac{4 \pi}{3}$ rotations
of $A_1$, respectively (see Fig.~\ref{fig:island}). 
Note that $A_1 \cap A_2 \cap A_3= \varnothing$.
Then, the intersection of the entanglement wedges of the three subregions
is an island-like region $M_{A_1} \cap M_{A_2} \cap M_{A_3}$, which does not reach the boundary, around the center of the disk as Fig.~\ref{fig:island}.
Now, let us consider a Hermitian bulk operator $\phi$ supported only on this island-like region.
The bulk operator $\phi$ can be reconstructed from
the CFT operator supported on any one of the subregions $A_i$, 
i.e. $\phi=\phi_i$ for 
$\phi_i \in {\cal A}(A_i)$ for $i=1,2,3$, where ${\cal A}(A_i)$ is the algebra generated by the CFT operator supported on $D(A_i)$.
If the (strong) entanglement wedge reconstruction holds, these operators are the same ($\phi_1=\phi_2=\phi_3$) on the code (low-energy) subspace \cite{Almheiri:2014lwa}. 
However, we will see that they are different even in the low-energy subspace.
We define\footnote{
Instead of $e^{i \phi}$, we can consider the Taylor expansion of it for some order like \eqref{psiG-e}, and the conclusion does not change.} 
the states $\ket{\psi_i}= e^{i \phi_i} \ket{\psi_0}$ for $i=1,2,3$
where $\ket{\psi_0}$ is an arbitrary state. 
They  exactly satisfy
\begin{align}
   \bra{\psi_i} {\cal O}_{\bar{i}} \ket{\psi_i}=\bra{\psi_0} {\cal O}_{\bar{i}} \ket{\psi_0},
   \label{vev1}
\end{align}
for arbitrary  
${\cal O}_{\bar{i}}\in D(\bar{A_i})$
by the causality in the CFT.
If we suppose $\phi_1=\phi_2=\phi_3$, these states are the same ($\ket{\psi_1}=\ket{\psi_2}=\ket{\psi_3} \equiv \ket{\psi}$). 
This implies 
\begin{align}
    \bra{\psi} {\cal O} \ket{\psi}=\bra{\psi_0} {\cal O} \ket{\psi_0}
    \label{vev2}
\end{align}
for any CFT operator ${\cal O}$ because
$\bar{A_1} \cup \bar{A_2} \cup \bar{A_3} $ is the entire space,\footnote{For the GFF theory, we have the additivity anomaly \cite{Casini:2019kex, Leutheusser:2022bgi}, i.e. ${\cal A}(A)\vee {\cal A}(A')\neq {\cal A}(A \cup A')$. 
Thus, ${\cal A}(A_1)\vee {\cal A}(A_2)\vee {\cal A}(A_3)\neq {\cal A}(S^1)$, and we cannot conclude $\bra{\psi} {\cal O} \ket{\psi}=\bra{\psi_0} {\cal O} \ket{\psi_0}$ for any operator ${\cal O} \in {\cal A}(S^1)$. Our discussion in the main text is for the finite $N$ CFT without the additivity anomaly. Although the entanglement wedge reconstruction may hold for the GFF theory, it does not contradict the radial locality due to the additivity anomaly, and thus we do not need the QEC structure in this $N=\infty$ case.} 
and then we conclude $\ket{\psi} \sim \ket{\psi_0}$ which means that $\phi$ is proportional to an identity operator.
Thus, there is no such non-trivial operator $\phi$ satisfying $\phi=\phi_1=\phi_2=\phi_3$.
Instead, these operators $\phi_1, \phi_2, \phi_3$ are different, that is the subregion complementarity.  
It is natural that there are no nontrivial operators satisfying \eqref{vev2}
because there is no overlap of the CFT subregions $A_i$, i.e. $A_1 \cap A_2 \cap A_3= \varnothing$.
This is a typical example of the subregion complementarity.

The statement that $\phi_1, \phi_2, \phi_3$ are different operators sounds trivial from the CFT perspective because they have different supports.
However, this is not trivial for low-energy subspace because we need clarification for the concept of the local operators in low-energy effective theories, as we will discuss in subsec.~\ref{subsec:bulk-algCFT}.
It will be clear that they are different from the bulk side when we take into account the gravitational dressing (see sec.~\ref{sec:gauge-inv}).

One might think that there exists some non-trivial bulk operator $\phi$
which satisfies 
$\phi \simeq \phi_i$ where
$\phi_i \in D(A_i)$ for $i=1,2,3$
only in the low-energy subspace (which is the code subspace) as the holographic quantum error correction proposal \cite{Almheiri:2014lwa}. 
However, 
one can repeat the above discussion by restricting 
$\ket{\psi_0}$ and 
${\cal O}_{\bar{i}}$
to a state and an operator in the low-energy subspace.
It is important to note that $\phi_i$ which is reconstructed by the CFT operators should be only supported on $A_i$
strictly and ${\cal O}_{\bar{i}}$ should be only supported on $\bar{A}_i$. 
This enables us to use the micro-causality.
Then,
we find that $\bra{\psi} {\cal O} \ket{\psi}=\bra{\psi_0} {\cal O} \ket{\psi_0}$ for any low-energy CFT operator $O$
and 
such $\phi$ should be proportional to the identity operator in the low-energy subspace.\footnote{
Strictly speaking, to show this, we need to assume that there is no genuine non-local low-energy CFT operator which 
cannot be supported on $A_j$ for any $j$.
However, such genuine non-local operators may not exist
because there may not be such low-energy excitations in the bulk.
Furthermore, even if we assume the existence of such a non-local operator, %${\cal O}_{\rm non-local}$,
$\phi_i$ should include this non-local operator, which cannot be supported on $A_j$ for any $j$.
This is impossible because $\phi_i$ is supported on $A_i$.
}
Thus, such an operator should be trivial in the low-energy subspace and it is contrary to the holographic quantum correction code proposal.

\paragraph{Generalization}
We can extend the discussion to general backgrounds and more general choices of subregions.  
Let us consider a time slice of an asymptotic AdS spacetime $M$ and
CFT subregions $A_i \subseteq \partial M$. 
We will take bulk regions 
$a_i \subseteq M$ such that
\begin{align}
    a_i \cap \partial M = A_i,
    \label{bew}
\end{align}
for any $i$. 
Though this condition is satisfied for the time-slice of the entanglement wedge of $A_i$, we here allow $ a_i$ to be more general bulk subregions satisfying this condition.
We also require 
the following condition,
\begin{align}
\cup_j \bar{A}_j =\partial M,
\label{cft1}
\end{align}
which is also satisfied with the previous example.
(We do not require $\cup_j A_j =\partial M$ here.)
The condition \eqref{cft1} is equivalent to 
\begin{align}
    \cap_j A_j =\varnothing,
\label{cft2}
\end{align}
which implies that there is no CFT operator $\phi$ such that $\phi \in {\cal A}(A_i)$ for all $i$, where ${\cal A}(A_i)$ is the algebra generated by the CFT operators supported on $A_i$.
Under the condition \eqref{bew},
\eqref{cft2}
is equivalent to
\begin{align}
    \cap_j a_j \cap \partial M =\varnothing,
    \label{bulk1}
\end{align}
which means 
that the intersection of the bulk wedges $a_i$ is an ``island'' which does not reach the boundary.
Thus, this can be a generalization of the previous example
because all of the conditions here are satisfied in the previous examples.
We can repeat the previous discussion and obtain the same conclusion for this generalized case. That is, 
if the overlapped region of the bulk wedges of the subregions is an island region and we reconstruct a bulk operator supported only on the island region as a CFT operator supported on a subregion $A_i$, then the CFT operator cannot be the same as the reconstruction from another subregion $A_j$ $(i\neq j)$ even in the low-energy subspace.

\section{Gravitational dressing (Gauge-invariant operators in gravity) and subregion complementarity}
\label{sec:gauge-inv}

We have seen that the CFT operator on a subregion that reconstructs a bulk ``local'' operator
is different from the other one for a different subregion
even restricted in the low-energy subspace.
This is because of the subregion complementarity.
We will explain that this property is naturally understood in the bulk gauge invariant description.
It is important that 
there are no gauge (diffeomorphism) invariant local operators in the gravitational theory,
and the gravitational dressing (or the gravitational Wilson lines) is needed to make a naive local operator gauge invariant \cite{Donnelly:2015hta, Donnelly:2016rvo}.\footnote{
For gauge theories also, such a dressing is required.
However, there are gauge-invariant local operators in
the gauge theories and the non-local operators may not 
play an important role in some situations, like 
the low energy physics of the large $N$ gauge theories because the Wilson loops will be heavy.
For the gravitational theory, there are only non-local gauge invariant operators. 
}
We will also show that subregion complementarity, rather than the holographic QEC proposal, is consistent with the algebraic approach to the entanglement wedge reconstruction based on 
the gauge-invariant operators in gravity.
Note that this version of the entanglement wedge reconstruction is 
the weak version of the entanglement wedge reconstruction, which we claimed in \cite{Terashima:2021klf, Sugishita:2022ldv}, and is different from the usual entanglement wedge reconstruction.

\subsection{Gravitational dressing}
The examples of the subregion complementarity given in the previous section can be understood from the fact that 
there are no local operators in the gravitational theory (without a gauge fixing).
In gauge theories, charged objects are always accompanied by electromagnetic fields. 
In other words, we have to attach the Wilson lines to the charged operators in order to make them gauge-invariant.
In the gravitational theory, gravitational fields are universally coupled with all fields, and thus all operators are accompanied by gravitational fields.
Thus, we have to attach the gravitational dressing or the ``Wilson line''\footnote{
The gravitational analogue of the Wilson line is different from the Wilson line in the gauge theories
because there is no analogue of the Wilson loop.
In particular, for gauge theories, the Wilson line can be deformed arbitrarily by 
adding the Wilson loop, which is gauge-invariant and localized inside the bulk. 
}  
to the local field to make it gauge (diffeomorphism) invariant \cite{Donnelly:2015hta, Donnelly:2016rvo}.
In the gauge theories, the Wilson lines from a charged particle do not have to end on the boundary because they can end on the anti-charged particles. 
The gravitational dressing has to end on the boundary of the bulk spacetime because there are no ``anti-charged'' objects for the gravitational force.

\begin{figure}[htbp]
%\vspace{-0.01\columnwidth}
\centering
\includegraphics[width=15em]{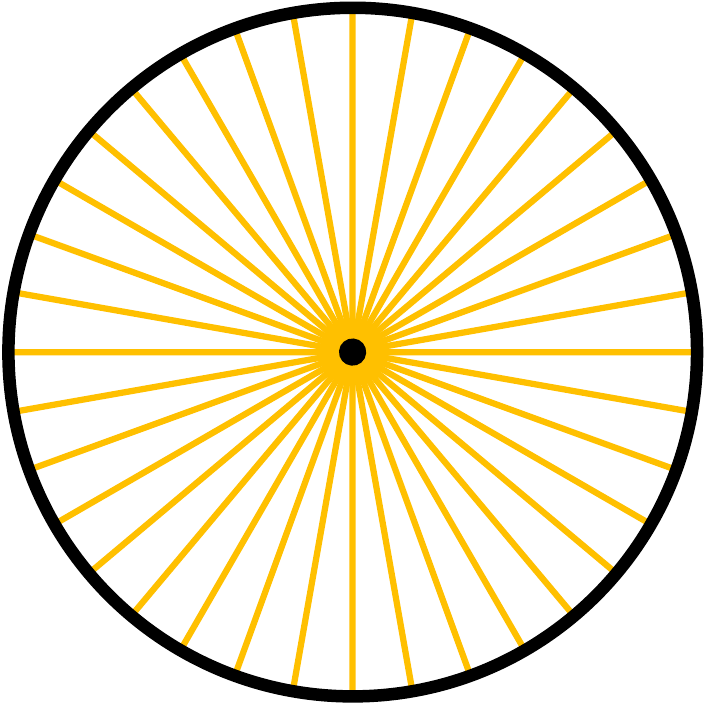}
\hspace{5em}
\includegraphics[width=15em]{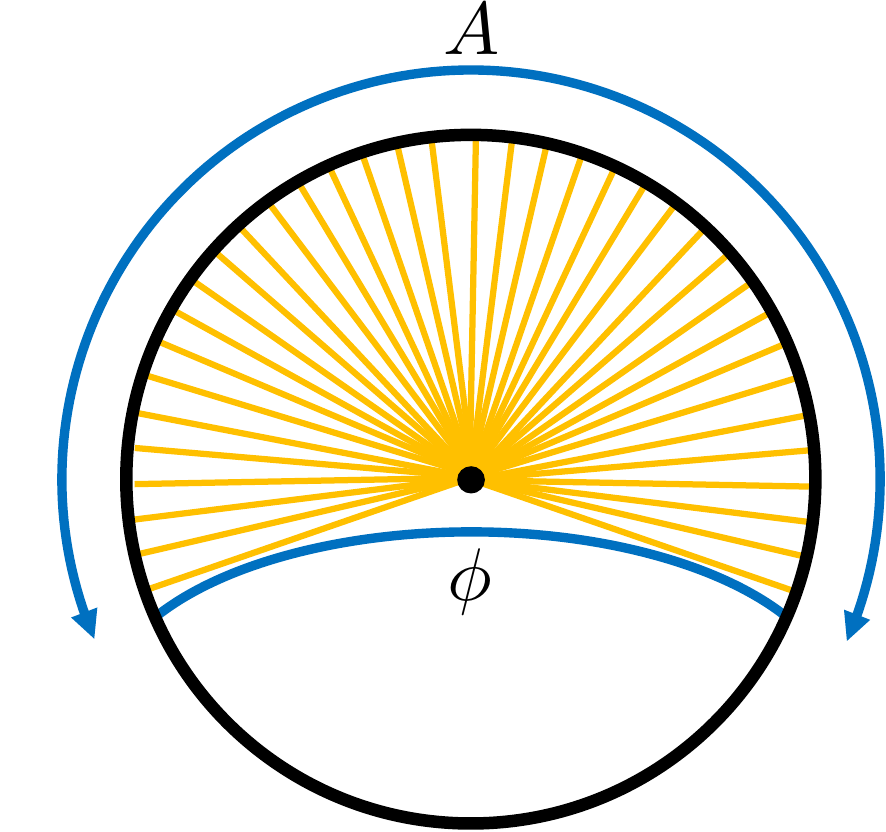}
\vspace{-0.01\columnwidth}
\caption{Operators with gravitational dressings. 
In the left figure, the bulk operator has the gravitational dressing that extends on the entire boundary. 
In the right figure, the gravitational dressing extends on the entire boundary. 
The two bulk operators are different for the finite $N$ theory.}
\label{fig:dress}
\vspace{-0.01\columnwidth}
\end{figure}

In the left figure in Fig.~\ref{fig:dress}, the bulk operator has the gravitational dressing which extends on the entire boundary. 
This operator corresponds to the CFT operator for the global reconstruction.
On the other hand, we can consider other choices of gravitational dressing as in the right figure in Fig.~\ref{fig:dress}. 
If the gravitational dress extends only on a subregion $A$ of the boundary, the operator may be constructed from the CFT operators supported only on $A$. 
The CFT operator is different from the one for the global reconstruction. 
The physical difference is clear because they have different gravitational fields.

%%%%%%%%
%%%%%%%%
%%%%%%%%
%%%%%%%%

In the example (Fig.~\ref{fig:island}) in the previous section, the operator $\phi_1$ constructed from $A_1$ corresponds to the bulk operator with the gravitational dressing extending (only) to $A_1$. 
Similarly, $\phi_2$ and $\phi_3$ correspond to the bulk operators with gravitational dressing extending (only) to $A_2$ and $A_3$, respectively. 
Therefore, the bulk operators including the gravitational dressing have different supports, and it is consistent with our observation that CFT operators $\phi_1, \phi_2, \phi_3$ are different. 

We have seen that the constraint of the gauge invariance of the bulk theory is related to the causality of the CFT side.
It may be interesting to understand such a relation clearly.

\paragraph{No operators supported on island}
Gravitational dressing must end on the boundary. 
Thus, there are no gauge invariant operators supported only on an island region in the bulk.
This is consistent with the examples in the previous section because there are no CFT operators associated with the island region.
This may imply that 
an island subregion in an entanglement wedge is irrelevant for 
the operators supported on the wedge.
Let us take
a certain bulk subregion $M_A$ which is connected to the bulk boundary. 
If we consider
another subregion $M'_A=M_A \cup I$ where
$I$ is an island subregion that is not connected to the bulk boundary and also $M_A$, 
then the bulk subalgebra of the operators supported on $M_A$ and 
the one on $M'_A$ are the same because there are no gauge-invariant operators localized on $I$.
Note that this conclusion is based only on the bulk consideration, and
the discussion here may be equivalent to the one in \cite{Geng:2021hlu} (see also \cite{Geng:2020qvw, Geng:2020fxl})
although the holographic error correction codes \cite{Dong:2016eik} and 
the (original version of) entanglement wedge reconstruction \cite{Almheiri:2014lwa} were supposed to be correct in \cite{Geng:2021hlu}.

\subsection{Subalgebra associated with subregion and the algebraic entanglement entropy}

The subregion duality and the entanglement wedge reconstruction \cite{Dong:2016eik} are based on the equivalence of the bulk and CFT relative entropies \cite{Jafferis:2015del}.
To define the (conventional) entanglement entropy or the reduced density matrix for a space subregion,
a tensor-factorized structure is needed for the Hilbert space.
However, without a gauge fixing, 
there are no local gauge invariant operators in
the gravitational theory and 
there is no such tensor-factorized structure.
Instead of a tensor-factorized structure associated with a space subregion,
we can consider a subalgebra
that is generated by operators supported on the subregion.
We have a generalization of the definition of entanglement entropy based on
the tensor-factorized structure to the one based on the subalgebra (see, e.g., \cite{Casini:2013rba, Radicevic:2016tlt, Harlow:2016vwg}), which is also reviewed in appendix~\ref{app:alg}.
Note that this generalized definition reduces to the conventional definition when the total Hilbert space is tensor-factorized as $\cH_1 \otimes \cH_2$ by considering the subalgebra non-trivially acting only on a factor of the tensor product, e.g., $\cH_1$.
Thus, the definition of the reduced density matrix based on a subalgebra may be a natural generalization.
Indeed, in \cite{Harlow:2016vwg, Cotler:2017erl},
the subregion duality and the entanglement wedge reconstruction based on the algebraic approach were proposed by 
(implicitly) assuming that the relative entropy computed in \cite{Jafferis:2015del} is based on the algebraic approach.
It is not yet known whether this assumption is correct
because 
the replica trick was used to obtain the density matrix for the subregion in \cite{Jafferis:2015del, Lewkowycz:2013nqa} and
the replica trick is based on the tensor-factorized structure of the Hilbert space at least naively.\footnote{In \cite{Ghosh:2015iwa},
it was argued that the entanglement entropy calculated using a definition of the replica trick will be the algebraic entanglement entropy for gauge theories.
}

We claim that the subregion duality and the entanglement wedge reconstruction based 
on the subalgebra are what we call the weak version of them \cite{Terashima:2021klf, Sugishita:2022ldv}.
This implies that the entanglement wedge reconstruction based on the tensor product of the Hilbert space is completely different from
those based on the subalgebra
although it seems to be assumed that they are not so different in \cite{Harlow:2016vwg, Cotler:2017erl}.
In particular, the subregion complementarity is relevant in our discussion instead of the holographic QEC proposal.

\subsection{What is the CFT counterpart of the bulk subalgebra}

Let us consider ball-shaped subregions $A, B$ in CFT and 
the corresponding AdS-Rindler patches $M_A, M_B$ in the bulk. We allow $A$ as the entire space $S^{d-1}$, i.e. $M_A$ as the global AdS space.
Then, we can reconstruct a bulk local operator $\phi(x)$
where $x \in M_A \cap M_B$ by either of the AdS-Rindler HKLL reconstruction from the subregions $A$ or $B$.
The reconstructed operators from $A$ and $B$ should be different as discussed in the previous section.
In the gauge invariant language, 
the bulk local operator $\phi(x)$ should be supplemented by the gravitational dressing.
The gravitational dressings also should be supported on $M_A$ or $M_B$ for the reconstructed operators from $A$ or $B$, respectively,
if the algebraic version of the subregion duality holds.
Then, the gravitational dressing is expected to  
give the nonzero three-point function of the CFT energy-momentum tensor and two $\cO$ corresponding to $\phi$ only on $A$ or $B$, respectively. Thus, the two reconstructed operators are different because they have different three-point functions.

The above claim can also be further explained with more physics-related implications by using the simple picture given in \cite{Terashima:2020uqu, Terashima:2021klf} which are based on 
the studies of the AdS/CFT in the operator formalism \cite{Terashima:2017gmc, Terashima:2019wed}.
The bulk local operator can be represented by the time evolution of the CFT primary operators on the intersection of the asymptotic AdS boundary and the lightcone of it.
In particular, the bulk wave packet operator can be 
represented by the time evolution of the CFT primary operators at the point where  
the trajectory of the wave packet, which is a null geodesic for the well-localized wave packet, intersects with the asymptotic AdS boundary \cite{Terashima:2023mcr, Terashima:2021klf}.
Note that the CFT primary operators are regarded as 
the bulk operators on the boundary which are gauge invariant because the gravitational dressings are not needed. 
Then, the time evolution of them in the bulk picture
produces the operator analogues of the gravitational waves (and other waves)
by the bulk interactions.
They spread within $M_A$ because of the bulk causality. The gravitational dressing of the bulk operator is localized only on $M_A$.
Then, the AdS-Rindler HKLL reconstruction of the bulk local operator from the subregion $A$ differs from the bulk local operator from the subregion $B$.
In particular, if we consider a wave packet operator that is of the boundary-to-horizon type \cite{Sugishita:2022ldv} for the AdS-Rindler patch $M_A$, but of the horizon-to-horizon type for the patch $M_B$, then the reconstructed operators from $A$ and $B$ should be different because the gravitational dressing of the former should extend beyond $M_B$.

\subsubsection{The bulk subalgebra in terms of CFT operators}\label{subsec:bulk-algCFT}

Here, we will express the bulk subalgebra in terms of CFT operators. As we will see below, we need a non-trivial clarification.

In the CFT picture, there is
the algebra ${\cal A}_{CFT}$ which is generated  
by the CFT operators.
Around a semiclassical bulk background, which corresponds to the vacuum for our case here,
we will define the low-energy Hilbert space, whose basis is spanned by the states with energy less than ${\cal O}(N^0)$, and the subspace may be called the code subspace.\footnote{
There are some subtleties to defining the low-energy subspace because the ${\cal O}(N^0)$ energy is only defined in the large $N$ limit and we need to introduce an explicit cutoff for the finite $N$ case. 
We will ignore this subtlety, assuming that the discussions below are not sensitive to the explicit cutoff.}
The operators of the semiclassical bulk theory will act on this.
We have considered 
the algebra generated ${\cal A}_{bulk}$ by the bulk operators.\footnote{
There are also some subtleties to define ${\cal A}_{bulk}$
because if the number of the products of low-energy operators or the number of derivatives is very large,
the corresponding operators cannot be regarded as low-energy ones.
We will ignore this, assuming the approximate notion of the subalgebra is enough.}
Note that ${\cal A}_{bulk}$ can be identified as a subalgebra of ${\cal A}_{CFT}$, by considering the operators acting on the low-energy states only.\footnote{If we consider another excited state $\ket{\psi}$ corresponding to the non-trivial semiclassical background, the low-energy states mean the states excited by acting low-energy operators on $\ket{\psi}$.
}
Here, the low-energy operators will be defined such that
the matrix elements between low-energy states (whose energy is equal to or less than $\cO(N^0)$) and high-energy states (whose energy is equal to or bigger than the Planck mass) vanish (or are small compared with $1/N^n$ where $n$ is an arbitrary positive integer).
More precisely, for the identification of ${\cal A}_{bulk}$ with a subalgebra in CFT, we need to 
extract
the matrix elements between the low-energy states.

Now, we consider a subregion $A$ in CFT,
and then we have the subalgebra ${\cal A}_{A}$ of 
${\cal A}_{CFT}$
which is generated by 
the CFT operators supported on $A$.\footnote{
We will ignore non-local CFT operators, which will be high-energy operators.
}
A central question of the entanglement wedge reconstruction might be which part of the bulk algebra ${\cal A}_{bulk}$ can be generated by ${\cal A}_{A}$,
in other words, what is the low-energy subalgebra of ${\cal A}_{A}$.
However, this is not a good question because 
the bulk theory is meaningful only for the low-energy subspace and operators involving multiple derivatives 
cause some problems
as discussed in \cite{Terashima:2020uqu}.
We will explain this below.

The bulk theory may have a UV cutoff at the Planck scale. Let us remember the subtlety of the operators with a large number of derivatives
in the cutoff theory
discussed in \cite{Terashima:2020uqu}.
For the notational simplicity, we will consider the CFT on the Minkowski space.
Let us consider the following operator
\begin{align}
    {\cal O}(\bar x, \bar t ) = e^{\bar{t} \partial_t} {\cal O}(\bar x, t ) |_{t=0}.
\end{align}
It is not localized at $\{ \bar x, t=0 \}$, but at $\{ \bar x, \bar t \}$.\footnote{
In the discussions here we will consider time derivatives for simplicity although we can generalize them to any direction. 
}
It means that if we act the infinite number of derivatives on the local operator at a point, the obtained operator is not local at the original point.
On the other hand, if the number of derivatives is finite as
\begin{align}
\label{def:Oq}
     {\cal O}^{[q]}(\bar x, \bar t ) \equiv  
     [e^{\bar{t} \partial_t}]_q {\cal O}(\bar x, t ) |_{t=0},
\end{align}
where 
\begin{align}
    [e^{\bar{t} \partial_t}]_q \equiv \sum_{n=0}^{q} \frac{(\bar{t} \partial_t)^n}{n!},
\end{align}
it is the local operator at $\{ \bar x, t=0 \}$ for a UV complete quantum field theory.

For a low-energy effective theory with the energy cutoff $\Lambda$,
the local operator should be smeared,
for example, as 
\begin{align}
     {\cal O}_{\Lambda}(x, t )\equiv \int dt' e^{-\frac{\Lambda^2}{2}(t-t')^2} {\cal O}(x, t' ).
\end{align}
Here, the cutoff $\Lambda$ may be restricted to be much smaller than the Planck mass $M_p \sim N^{\frac{2}{d-1}}$ for the bulk theory.
Then, 
the ``local'' operator 
$ {\cal O}_{\Lambda}(\bar{x}, \bar{t} )$ in the effective theory
cannot be distinguished from
\begin{align}
     {\cal O}^{[q]}_\Lambda (\bar x, \bar t ) \equiv  
     [e^{\bar{t} \partial_t}]_q \, {\cal O}_\Lambda (\bar x, t ) |_{t=0}
\end{align}
if $q \gg \bar{t}  \Lambda$, 
because
\begin{align}
     \frac{(\bar{t} \partial_t)^q}{q!} 
     \sim e^{q(\ln (\bar{t} \partial_t)-\ln q )},
\end{align}
and the Fourier mode of ${\cal O}_\Lambda (\bar x, t ) $ for $t$ is exponentially suppressed for $\omega \gg \Lambda$.
Note that if we take $q=M_p ( \lesssim N^2) $ the condition $q \gg \bar{t}  \Lambda$ is satisfied for $\bar{t}={\cal O} (N^0)$ and the difference ${\cal O}^{[q]}_\Lambda (\bar x, \bar t )- {\cal O}_{\Lambda}(\bar{x}, \bar{t} )$ is suppressed by a factor $e^{-q}$ at least.
Thus, as the effective theory, 
such higher derivative terms cannot be considered to be a local operator at $t=0$ although 
it is a local operator at $t=0$ for the UV complete CFT.\footnote{
This can be explicitly seen in the lattice field theory.
}
For the AdS/CFT correspondence, the important thing is that
the large $N$ gauge theory has ${\cal O} (N^2)$ degrees of freedom  and 
$ (\partial_t)^q {\cal O}( x, t )$ 
with $q \lesssim N^2$ are independent fields
by imposing the equations of motion.
Even though this fact, ${\cal O}^{[q]}_\Lambda (\bar x, \bar t )$ with $q ={\cal O} (N^2)$ cannot be regarded as a local operator at $t=0$
for the bulk effective theory.

Thus, when we take a subregion $A$ and consider CFT operators on $A$ with a large but finite number of derivatives like ${\cal O}^{[q]}(\bar x, \bar t)$ in \eqref{def:Oq},  these operators are not local operators on $A$ in the EFT, and can be local operators on other subregions.
It means that the CFT operators supported only on a subregion can be operators supported on any subregion in the EFT (which corresponds to the bulk description )using a huge number of derivatives, and thus they can be bulk operators anywhere (because of the UV cutoff).
This leads that instead of ${\cal A}_{A}$,  
we need to consider ${\cal A}_{A}^{\Lambda}$
which is generated by CFT operators without a huge number ($N^{\frac{2}{d-1}}$) of derivatives supported on $A$.\footnote{
From the viewpoint of the CFT living on the subregion $A$, such a restriction may come from the coordinate choice.
In the Rindler coordinate, the boundary is at the infinite limit of the coordinates.
More precisely, the boundary condition on the boundary of $A$ is important and we employed it implicitly by taking the Rindler coordinate.
}
Then, we find that 
${\cal A}_{A}^{\Lambda}$ will be generated by 
\begin{align}
    \int_{D(A)} K(X) {\cal O}_a(X),
    \label{ALA}
\end{align}
where $K(X)$ is a smooth function such that
the Fourier modes of it above the cutoff $\Lambda$ are suppressed 
and ${\cal O}_a$ is any low-energy single trace primary operator.
Note that we need CFT operators supported on $D(A)$ instead of $A$ because some independent CFT operators, which have a huge number of derivatives, on $A$ are not included in ${\cal A}_{A}^{\Lambda}$.\footnote{
CFT primary operators at different time correspond to
bulk operators at $t=0$ with different values for the radial coordinate.
}
To compare this to the bulk subalgebra, which acts on the low energy subspace only, we define ${\cal A}_{A}^{low}$ as the restriction of 
${\cal A}_{A}^{\Lambda}$ to the low energy subspace.\footnote{
It is important to note that the low energy part of ${\cal A}_{A}$, instead of that of ${\cal A}_{A}^{\Lambda}$, was supposed to be dual to the bulk subalgebra in the literature.
However, it is not correct as we have explained.
}

We can show that ${\cal A}_{A}^{low}={\cal A}_{M_A}$ for the ball-shaped subregion $A$ in CFT whose entanglement wedge $M_A$ is the causal wedge of $A$ covered by the AdS-Rindler patch as follows. 
Here, the subalgebra ${\cal A}_{M_A}$ is generated by
bulk gauge-invariant operators, including the gravitational dressings, supported on $M_A$.
First, we can easily see that ${\cal A}_{A}^{low} \subseteq {\cal A}_{M_A}$ because the CFT operator ${\cal O}_a(X)$ in \eqref{ALA} is localized on $D(A)$, which lies on the asymptotic boundary in the bulk picture. By the bulk equations of motion and causality, it can only affect the causal wedge $M_A$.
Next, let us consider ${\cal O}^{bulk} \in {\cal A}_{M_A}$ and represent it using low-energy CFT primary operators. Recall that the gravitational dressing in ${\cal O}^{bulk}$ is also localized on $M_A$.
To represent it in CFT, we do not need CFT primary operators outside $D(A)$, because such operators cannot be supported on $D(M_A)$ in the bulk picture.
Furthermore, a CFT operator supported on $D(A)$ with a large number of derivatives will be a high-energy operator unless it is an operator at a different point approximately. 
Therefore, we find that ${\cal A}_{A}^{low} = {\cal A}_{M_A}$.
Thus, the simple picture presented in \cite{Terashima:2020uqu} and \cite{Terashima:2021klf}
is consistent with the algebraic version of the JLMS \cite{Jafferis:2015del} and the entanglement wedge reconstruction based on it \cite{Harlow:2016vwg, Cotler:2017erl}.

\paragraph{Hawking radiation and information paradox}

What we have shown implies that for the information theoretical aspect of the AdS/CFT or quantum gravity,
the bulk local operator or state cannot be an appropriate approximation
and we need to consider some subregion including the asymptotic boundary.
This is the key point of the subregion complementarity.
Thus, if we want to consider, for example, the Unitarity of the Hawking radiation process for the information paradox, 
we need to specify what the Hawking radiation is in this sense, because the derivation of the Hawking radiation uses the semiclassical approximation, which does not specify the associated gravitational dressings. 
This is related to the holography of information, which was recently discussed in \cite{Laddha:2020kvp, Chowdhury:2020hse}.
In the holography of information, the important fact is that there is a difference between a non-gravitational theory and a gravitational theory concerning information.
We can say that what is important in our paper is that  
the difference between a $N=\infty$ (or its $1/N$ perturbation theory) theory and a finite $N$ theory concerning information.
Here, a $N=\infty$ theory corresponds to a non-interacting bulk gravity theory,
which is similar to a non-gravitational theory in a sense.
Thus, our discussions may be related to the holography of information.

\subsection{Simple derivation of no holographic quantum correction code structures for the bulk reconstruction}

Here, we argue that there are no holographic quantum correction code structures for the bulk reconstruction.
More precisely, we claim that if we can reconstruct a bulk operator as an operator ${\cal O}^A$, which is a low energy operator in the CFT, supported on subregion $A$ in CFT, as well as an operator ${\cal O}^B$ supported on subregion $B$,
then, if they are the same operator at least in the low energy subspace, we can reconstruct it as an operator supported on subregion $A \cap B$ in CFT as follows.

First, we assume that such operators ${\cal O}^A, {\cal O}^B$ are the same in the low energy subspace.
The operator ${\cal O}^A$ supported on subregion $A$ should commute with the low-energy operators, including the energy-momentum tensors, supported on $D(\bar{A})$ because of the micro causality.\footnote{More precisely, we need to smear the energy-momentum tensors in order to make them low-energy operators. ${\cal O}^A$ and ${\cal O}^B$ are not local operators and are smeared appropriately. 
For non-trivial CFTs, such smearing should be in $D(A)$, not $A$ as shown in \cite{Nagano:2021tbu}. Therefore, strictly speaking, the support of the operator should be written as $D(A)$, but for simplicity of notation, we denote it as $A$.}
Then, ${\cal O}^B$ should also commute with them in the low energy subspace.
The ${\cal O}^B$ is a sum of products of local operators and we can decompose it to
${\cal O}^B ={\cal O}^B_{A \cap B} +{\cal O}^B_{\bar{A}}$ where ${\cal O}^B_{A \cap B}$ contains only the operators supported on $A \cap B$ and each term of ${\cal O}^B_{\bar{A}}$ always contain operators supported on $\bar{A} \cap B$.
Then, ${\cal O}^B_{\bar{A}}$ should
commute with the energy-momentum tensors supported on $\bar{A}$ in the low energy subspace.
This implies ${\cal O}^B_{\bar{A}}=0$ in the low energy subspace because 
the energy-momentum tensors are low energy operators and also the commutators of them and a local operator are essentially proportional to the local operator.
Note that 
such a commutator with the energy-momentum tensor does not cancel between each term, because the energy of a CFT operator is never negative.
Thus, we can redefine the reconstructed operator supported on $B$ by eliminating the part.
In other words, we have ${\cal O}^B_{A \cap B}$
as a reconstructed operator
instead of ${\cal O}^B$.
Thus, the bulk operator can be reconstructed as an operator supported on the subregion $A \cap B$ in CFT. 

The holographic quantum correction code proposal \cite{Almheiri:2014lwa} claims the existence of operators ${\cal O}^A, {\cal O}^B$ such that 
they are identical in the low energy subspace (code subspace) but their supports are not restricted to $A \cap B$.
% that there exist such operators ${\cal O}^A, {\cal O}^B$ that cannot be operators supported on subregion $A \cap B$,
% and thus,
Based on the above argument, we conclude that there is no holographic quantum correction code structure for the bulk reconstruction.

%\section{Discussions}

\section*{Acknowledgement}

%The author would like to thank XXXX for collaboration in the early stages of this work
%and helpful discussions.
The authors thank Hiroki Kanda, Taishi Kawamoto, Juan Maldacena, Yoshinori Matsuo, Yu-ki Suzuki, Yusuke Taki, Tadashi Takayanagi and Zhenbin Yang for the useful comments.
This work was supported by MEXT-JSPS Grant-in-Aid for Transformative Research Areas (A) ``Extreme Universe'', No. 21H05184.
This work was supported by JSPS KAKENHI Grant Number 	24K07048.
SS acknowledges support from JSPS KAKENHI
Grant Numbers JP21K13927 and JP22H05115.
\hspace{1cm}

%Note added:
%As this paper was being completed, we became aware of the preprint \cite{Kinoshita:2023hgc} in which

\appendix
\section{Algebraic definition of reduced density matrix and entanglement entropy}\label{app:alg}

When we define the reduced density matrix, we usually suppose a tensor factorized structure of the Hilbert space as $\cH=\cH_A \otimes \cH_{\bar{A}}$.
We then define the reduced density matrix $\rho_A$ on $\cH_A$ by taking the partial trace of a given total density matrix $\rho$ over $\cH_{\bar{A}}$ as $\rho_A= \tr_{\bar{A}} \rho$.

However, we encounter some problems when we try defining a reduced density matrix associated with a subregion in QFTs.
One issue is the UV problem. 
In QFTs, the reduced density matrix of a subregion is inherently ill-defined due to UV problems, necessitating some UV regularization. 
This is related to the fact that the subalgebra associated with a subregion in QFTs is type III. 
Here, we assume a UV regulator that circumvents this issue, such as in lattice field theories. 

Even setting aside the UV problem, 
there is another issue for gauge theories. 
The Hilbert space of gauge-invariant physical states is not a tensor product like $\cH \neq \cH_A \otimes \cH_{\bar{A}}$ concerning the subregion $A$ and its complement $\bar{A}$ due to the Gauss law constraint. 
It corresponds to the fact that the subalgebra for the subregion $A$ is not a factor for gauge theories due to the existence of a non-trivial center.\footnote{von Neumann algebra $\mathcal{A}$ is called a factor if the center $\mathcal{Z}=\mathcal{A} \cap \mathcal{A}'$ consists only of multiples of the identity operator, where $\mathcal{A}'$ is a commutant of $\mathcal{A}$. }
However, by adopting an algebraic approach (see, e.g.,  \cite{Casini:2013rba, Radicevic:2016tlt, Ghosh:2015iwa, Soni:2015yga, Harlow:2016vwg}), we can define the reduced density matrix for subregion $A$ even for this case.
This approach is also used to define the target space entanglement entropy (see, e.g., \cite{Mazenc:2019ety, Das:2020jhy, Sugishita:2021vih}).

Let $\mathcal{A}$ be a subalgebra associated with the subregion $A$.
We define the reduced density matrix $\rho_{\mathcal{A}}$ associated with a subalgebra $\mathcal{A}$ from a given total density matrix $\rho$ as a positive semi-definite operator in $\mathcal{A}$ satisfying
\begin{align}
    \tr (\rho_{\cA} \cO)=\tr (\rho \cO)
\end{align}
for any $\cO \in \cA$.
In this definition, we do not need a tensor factorized form of operators in $\mathcal{A}$ like $\cO_A \otimes 1_{\bar{A}}$.
Nevertheless, by taking an appropriate basis, the total Hilbert space $\cH$ can be decomposed into a direct sum of tensor products as\footnote{See, e.g., \cite{Harlow:2016vwg}. Here, as discussed above, we assume that a UV regulator is introduced and also that the total space is compact such that the total algebra is type I.}
\begin{align}
    \cH = \bigoplus_k \cH_A^{(k)}\otimes \cH_{\bar{A}}^{(k)}
\end{align}
such that subalgebra $\cA$ takes a tensor-factorized form in each sector as
\begin{align}
    \cA=\bigoplus_k \mathcal{L}(\cH_A^{(k)}) \otimes 1_{\cH_{\bar{A}}^{(k)}}
\end{align}
where $\mathcal{L}(\cH_A^{(k)})$ denotes a set of operators on $\cH_A^{(k)}$.
The subalgebra $\cA$ does not mix different sectors labeled by $k$,\footnote{
The label $k$ might be continuous parameters. 
In this case, the direct sum is replaced by the direct integral.
}
and thus $k$ is the label of the superselection sectors for observables in $\cA$.

Let $\Pi^{(k)}$ be the projection from $\cH$ onto the $k$-sector $\cH^{(k)}:= \cH_A^{(k)}\otimes \cH_{\bar{A}}^{(k)}$. 
Then, from the total density matrix $\rho$, we can define the density matrix on the $k$-sector by using the projection as
\begin{align}
    \rho^{(k)}:= \frac{1}{p^{(k)}} \Pi^{(k)} \rho \Pi^{(k)},
\end{align}
where $p^{(k)}:=\tr (\Pi^{(k)} \rho \Pi^{(k)})$ is a normalization factor so that $\rho^{(k)}$ is a normalized density matrix on $\cH^{(k)}$ as $\tr \rho^{(k)}=1$. 
We can regard $p^{(k)}$ as the probability of finding the given state $\rho$ in the $k$-sector. 
Since $\cH^{(k)}$ takes a tensor factorized form as $\cH^{(k)}= \cH_A^{(k)}\otimes \cH_{\bar{A}}^{(k)}$, we can take the partial trace of $\rho^{(k)}$ with respect to $\cH_{\bar{A}}^{(k)}$ as
\begin{align}
    \rho^{(k)}_A:=\tr_{\cH_{\bar{A}}^{(k)}} \rho^{(k)}.
\end{align}
We then define the reduced density matrix on $A$ as
\begin{align}
    \rho_A:= \bigoplus_k p_k \rho^{(k)}_A.
\end{align}
Note that $\rho_A$ is a density matrix on the space $\cH_A:=\bigoplus_k \cH_A^{(k)}$ and normalized as $\tr_{\cH_A}\rho_A=\sum_k p_k =1$.
Entanglement entropy for $A$ is given by the von Neumann entropy of $\rho_A$ as
\begin{align}
    S_A(\rho):=-\tr_{\cH_A}\left(\rho_A \log \rho_A \right)
    =-\sum_k p_k \log p_k +\sum_k p_k S_{A}^{(k)}(\rho^{(k)}),
    \label{algSA}
\end{align}
where $S_{A}^{(k)}(\rho^{(k)})$ is given by
\begin{align}
    S_{A}^{(k)}(\rho^{(k)})=-\tr_{\cH_A^{(k)}}\left(\rho^{(k)}_A\log \rho^{(k)}_A \right),
\end{align}
that is, it is the entanglement entropy of $\rho^{(k)}$ for $\cH_A^{(k)}$ defined in the standard way.
If $\cH$ is tensor factorized as $\cH = \bigoplus_k \cH_A^{(k)}\otimes \cH_{\bar{A}}^{(k)}$, the above definition of entanglement entropy reduces to the standard one because in that case we have only a single sector (say $k=1$) and thus $p_{k=1}=1$.

We have a diffeomorphism symmetry in the bulk description. 
Thus, there are no gauge-invariant (diffeomorphism-invariant) local operators, and all operators have non-local gravitational dressing.
For bulk subregion $M_A$, we define the subalgebra $\cA_{M_A}$ associated with $M_A$ as the set of (low-energy) bulk operators that are supported only on $M_A$ including their gravitational dressing as Fig.~\ref{fig:dress}.
Due to this construction, the bulk subregion $M_A$ must have an asymptotic boundary where the gravitational dressing ends. 
For this subalgebra $\cA_{M_A}$, the above procedure defines the reduced density matrix $\rho_{M_A}$ and the entanglement entropy $S_{M_A}$. 

Let us consider an entanglement wedge $M_A$ associated with a boundary subregion $A$. 
At least if $A$ is a connected region,  
we expect that the above algebra $\cA_{M_A}$ should agree with the CFT (low-energy) subalgebra $\cA_{A}^{low}$ on $A$. 
For instance, 
when $M_A$ is the AdS-Rindler patch, we have $\cA_{A}^{low}=\cA_{M_A}$ as will be argued in subsec.~\ref{subsec:bulk-algCFT}, although the definition of low-energy subalgebra $\cA_{A}^{low}$ is not-trivial (see subsec.~\ref{subsec:bulk-algCFT}). 
Then, by construction, the reduced density matrix $\rho_A$ defined by the subalgebra $\cA_{A}^{low}$ is the same as $\rho_{M_A}$. 
We thus have the JLMS formula of the relative entropies \cite{Jafferis:2015del} in the weak sense \cite{Sugishita:2022ldv, Sugishita:2023wjm}.

The entanglement entropy $S_{M_A}$ has the classical Shannon entropy term as the first term of the RHS of \eqref{algSA}. 
This term is owing to the center of $\cA_{M_A}$ and may be related to the degrees of freedom localized on the RT surface $\Sigma_A$ as argued in \cite{Donnelly:2016auv, Harlow:2016vwg, Lin:2017uzr, Speranza:2017gxd, Camps:2018wjf, Takayanagi:2019tvn}. 
It is thus natural to expect the entropy to be related to local quantities on $\Sigma_A$ such as the area.

\section{Uniqueness of Reduced Unitaries}
\label{reducedtoA}

In this appendix, we consider $\mathcal H=\mathcal H_A\otimes\mathcal H_{\bar A}$ with
$d_{\bar A}:=\dim\mathcal H_{\bar A}$
and two unitary operators
\[
  U_1 = U_A\otimes\mathbbm 1_{\bar A},
  \qquad
  U_2\in\mathcal U(\mathcal H).
\]
Here, $U_1$ and $U_2$ are associated with $\tilde{\phi}^R$ and $\tilde{\phi}_G$ in section \ref{differentop}.
In the following, we will show that if their partial traces over $\bar A$ coincide, i.e. 
\[
  \operatorname{Tr}_{\bar A}(U_2)
    \;=\;
  \operatorname{Tr}_{\bar A}(U_1),
\]
then $U_2 = U_1$. This implies that, if $\tilde{\phi}^R \neq \tilde{\phi}_G$ in the whole Hilbert space, they should be different operators even restricted in $H_A$. 

First, define the difference
\[
  K \;:=\; U_2 - U_A\otimes\mathbbm 1_{\bar A}\,.
\]
By the assumption,
\begin{equation}\label{eq:trace-zero}
  \operatorname{Tr}_{\bar A}K=0, \quad \operatorname{Tr}_{\bar A}K^\dagger=0.
\end{equation}
Because $U_2$ is unitary,
\[
  \mathbbm 1
  = U_2^\dagger U_2
  = (U_A^\dagger\!\otimes\!\mathbbm 1_{\bar A} + K^\dagger)
    (U_A\!\otimes\!\mathbbm 1_{\bar A} + K)
  = \mathbbm 1
    + U_A^\dagger\!\otimes\!\mathbbm 1_{\bar A}\,K
    + K^\dagger U_A\!\otimes\!\mathbbm 1_{\bar A}
    + K^\dagger K.
\]
Hence
\begin{equation}\label{eq:unitary-expansion}
  U_A^\dagger\!\otimes\!\mathbbm 1_{\bar A}\,K
  + K^\dagger U_A\!\otimes\!\mathbbm 1_{\bar A}
  + K^\dagger K
  \;=\; 0.
\end{equation}
Taking $\operatorname{Tr}_{\bar A}$ of \eqref{eq:unitary-expansion}
and using \eqref{eq:trace-zero} gives
\[
  \operatorname{Tr}_{\bar A}(K^\dagger K)=0.
\]
Because $K^\dagger K$ is positive semi-definite, its partial trace is
zero only if $K^\dagger K=0$, hence $K=0$.
Therefore $U_2 = U_1$, completing the proof.

% \section{Coherent state}
% \label{a1}

\newpage 

%%%%%%%%%%%%%%%%%%%%%%%%%%%%%%%%%%%%%%%%%%%%%%%%%%%%%%%%%%%%%%
%%%%%%%%%%%%%%%%%%%%%%%%%%%%%%%%%%%%%%%%%%%%%%%%%%%%%%%%%%%%%%
\bibliographystyle{utphys}
\bibliography{ref-AdSCFT.bib}
%%%%%%%%%%%%%%%%%%%%%%%%%%%%%%%%%%%%%%%%%%%%%%%%%%%%%%%%%%%%%%
\end{document}